\documentclass[twocolumn]{aastex62}

\usepackage{soul}

\accepted{ApJ}

\shorttitle{Estimation of Fuel Efficiency}
\shortauthors{Sung \& Kwak}

\begin{document}

\title{Estimating the Fuel Supply Rate on the Galactic Disk from High Velocity Clouds (HVCs) Infall}

\correspondingauthor{Kyujin Kwak}
\email{kkwak@unist.ac.kr}

\author{Kwang Hyun Sung}
\affiliation{Physics Department, Ulsan National Institute of Science and Technology, Ulsan 44919, Republic of Korea}

\author{Kyujin Kwak}
\affil{Physics Department, Ulsan National Institute of Science and Technology, Ulsan 44919, Republic of Korea}


\begin{abstract}

Previous studies suggest that the estimated maximum accretion rate from approaching high velocity clouds (HVCs) on the Galactic disk can be up to $\sim 0.4 ~M_{\sun}~ yr^{-1}$. In this study, we point out that the hydrodynamic interaction between the HVCs and the Galactic disk is not considered in the traditional method of estimating the infall rate and therefore the true supply rate of fuel from HVCs can be different from the suggested value depending on the physical configurations of HVCs including density, velocity, and distance. We choose 11 HVC complexes and construct 4 different infall models in our simulations to give an idea of how the fuel supply rate could be different from the traditional infall rate. Our simulation results show that the fuel supply rate from HVC infall is overestimated in the traditional method and can be lowered by a factor of $\sim 0.072$ when the hydrodynamic interaction of the HVC complexes and the disk is considered. 

\end{abstract}



\section{Introduction} \label{sec:intro}
First discovered by Muller, Oort, and Raimond in 1963, High Velocity Clouds (HVCs) are known as neutral atomic hydrogen clouds moving with a velocity that deviates from the galactic rotation by up to $\sim 300 ~ km~ s^{-1}$. The typical size of an HVC complex can be a few to 15 kpc across with the H{\small I} mass in the range of $10^{5} \sim 5 \times 10^{6} M_{\sun}$. While the exact distances to the HVC complexes are yet questionable and should be further constrained, it is widely accepted at the current stage that a majority of the complexes exist at a distance $\leq$ 10 kpc \citep{thom2008,wakker2007,wakker2008,putman2012}. Suggested in recent studies is that HVCs have multiple origins including stripped gas material of nearby (dwarf) galaxies like the Magellanic Stream, Galactic fountains, and inflowing intergalactic gas \citep{wvw1997,blitz1999,wvw2013}. Assuming that the HVCs are inflowing intergalactic gas is often favored in theoretical Galactic Chemical Evolution (GCE) models, especially when it is metal-poor material that is supplied into the galaxy from HVCs. For example, in the GCE model with two main infall episodes, the present infall rate of primordial material is predicted as $\sim 0.4~ M_{\sun} ~ yr^{-1}$   \citep{chiappini2001,chiappini2008}\footnote{This two episode model with the infalling low-metal material was able to explain observational constraints in the solar vicinity including the ``G-dwarf problem".}. Note that the continual and at the same time occasional infall of HVCs can be a source of metal-poor gas with the mass accretion rate being up to $\sim 0.4~ M_{\sun}~ yr^{-1}$ although only about half of the currently infalling material has low-metallicity \citep{putman2012}. 

The infall rate is derived from the physical properties such as mass, velocity, and distance of the HVCs and traditionally given in the form of an equation given as,
\begin{equation}
\dot{M} = \sum \frac{M_{i}{v_i}}{D_i}~,
\label{eq:1}
\end{equation}
where $M_{i}$ is the mass of an HVC, $v_{i}$ is the observed radial velocity, and $D_{i}$ is the distance \citep{wakker2007,thom2008,putman2012}. The advantage of estimating the gas infall rate from the equation above is the simplicity that allows us to directly utilize the observed physical properties of HVC complexes. However, there are some obvious limitations of this approach due to the fact that $D_{i}$ will not be the same distance as the distance from the HVC to the point where it will collide with the galactic disk. Furthermore, $v_{i}$ will also be different from the true radial velocity with the possibility of acceleration/deceleration. But more significantly, we believe there is a different part in this method that can be reconditioned to better estimate the supply rate of material into the galactic disk from such gas inflow. The essence of the traditional method of infall rate estimation is the assumption of steady and full accretion of HVC mass until ``time to impact (=$v_{i}/D_{i}$)". Kinematic consequences that occur from the hydrodynamic interaction between the disk and HVCs are neglected in the traditional approach and therefore, in this study, we make an attempt to qualitatively show that the true fuel supply rate can change depending on the physical properties of the HVC regardless of the total infall rate itself. Further, we select 11 different HVC complexes and set up four different infall cases in our numerical simulations to give an idea how the fuel supply rate could be different from the traditional infall rate depending on the physical configuration of each HVC complex. 

\section{Simulation Methods} \label{sec:simmethods}
\subsection{Simulation Setup}
We use FLASH 2.5 \citep{fryxell2000} for our simulations, which is a modular, adaptive-mesh, parallel simulation code capable of handling general compressible flow problems. Message-Passing Interface (MPI) library is used for parallelization and the PARAMESH library manages Adaptive Mesh Refinement (AMR). The computation domain is configured in a 2-D cylindrical geometry that can extend up to 10 kpc in the horizontal (i.e., r-) direction and 40 kpc in the vertical (i.e., z-) direction depending on the distance and radius of the infalling HVC. As the HVC complex approaches and interacts with the galactic disk, the refinement level gradually increases to its maximum which corresponds to 9.7 pc $\times$ 9.7 pc spatial resolution for each cell. For the boundary conditions, it is reflecting at the left vertical axis (r = 0) and outflowing at the rest of the three remaining boundaries. The thickness of the gaseous disk is 250 pc \citep{rougoor1964} with the H{\small I} volume number density of 0.1 $cm^{-3}$ \citep{kalded2008}. The H{\small I} volume number density of the ISM is $1.0 \times 10^{-4} ~ cm^{-3}$.

\begin{table}[t!]
\caption{HVC Infall Parameters} \label{tab:hvcinput}
\tablenum{1}
\begin{tabular}{cccccc}
\tablewidth{0pt}
\hline
\hline
Name & $b$\tablenotemark{a} & $v_{dev}$\tablenotemark{b} & $\Omega$\tablenotemark{c} & $D$\tablenotemark{d} & $M$\tablenotemark{e} \\
 & [$^{\circ}$] & [$km~s^{-1}$] & [$^{\circ2}$] & [kpc] & [$10^{6}M_{\sun}$] \\
\hline
A & +40 & 141 & 288 & 8.0 & 1.0 \\
ACHV & -30 & 100 & 397 & 10.0 & 1.0 \\
ACVHV & -30 & 221 & 338 & 10.0 & 1.0 \\
C & +58 & 122 & 1546 & 10.0 & 5.0 \\
GCN & -31 & 243 & 130 & 20.0 & 0.22 \\
M & +62 & 95 & 174 & 4.0 & 1.0 \\
Smith & +13.4 & 73\tablenotemark{*} & 58 & 12.4 & 1.0 \\
WA & +32 & 126 & 102 & 8.0 & 0.13 \\
WB & +32 & 52 & 289 & 8.0 & 0.70 \\
WD & +32 & 106 & 253 & 4.4 & 1.1 \\
WE & -20 & 106 & 51 & 9.4 & 0.12 \\
\hline
\end{tabular}
\tablecomments{$^{a}$Galactic latitude, $^{b}$deviation velocity, $^{c}$solid angle, $^{d}$distance, and $^{e}$mass. The $^{*}$vertical velocity (i.e., $v_z$) of the Smith cloud is 73 $km~ s^{-1}$ for the intermediate case and 99 $km~ s^{-1}$ for the extreme case \citep{lockman2008}.}
\end{table}

\begin{figure*}[t!]
	\centering
    \includegraphics[width=2.0\columnwidth]{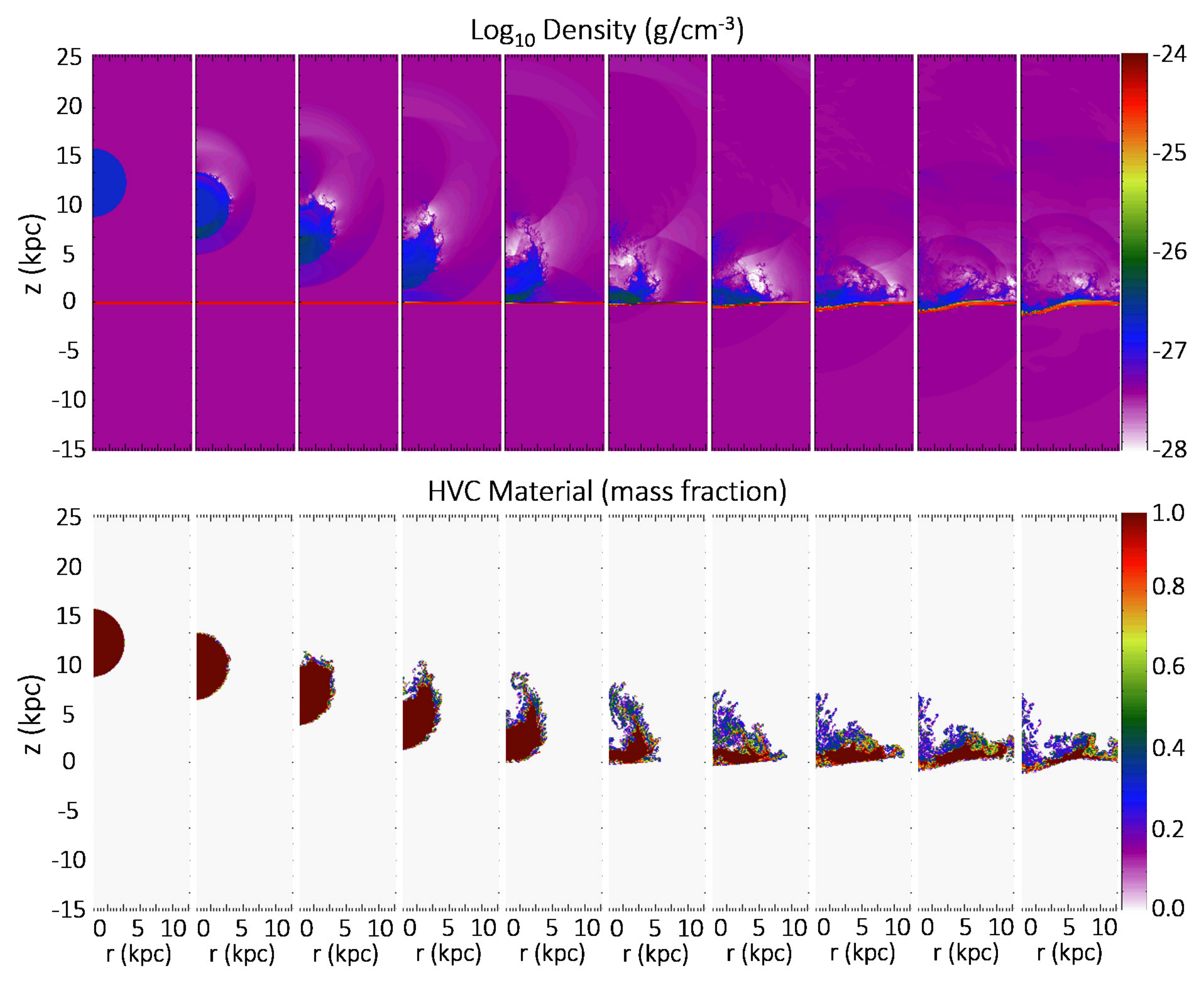}
    \caption{{\bf Time evolution of the collision between the disk and Complex C in terms of density (top) and HVC material (bottom). The panels are plotted at 20 Myr intervals for a time span of 180 Myr. The semicircles in the far left panels illustrate the HVC complex, the red horizontal layer in the density plots (top) is the gaseous disk of our galaxy, and the halo medium is present in the background. The complex with a radius of 3.43 kpc is falling down towards the disk with an initial velocity of 143.9 $km~ s^{-1}$ at the distance of 12.04 kpc from the disk to the center of the complex.}}
    \label{fig:1}
\end{figure*}

\subsection{HVC Infall Parameters} \label{subsec:hvcparam}
We select 11 HVC complexes from the HVC catalogue of \citet{wakker2005} and mostly adopt the provided information for the mass, distance, latitude ($b$), and deviation velocity ($v_{dev}$) but also  supplement the data set with observation results from recent studies \citep{thom2006,thom2008,lockman2008,jin2010,peek2016}. The input parameters for each complex are provided in Table 1. For each HVC complex, we consider two different cases for the vertical velocity, $v_z$; (1) assuming that the observed velocity is purely radial, $v_z = v_{dev} \sin(b)$ which we call intermediate case or (2) assuming that the space velocity is vertical, $v_z = v_{dev} / \sin(b)$ which we call extreme case. The radius, $R$ is estimated by assuming that the HVC complex is a perfect sphere and has a uniform density. Then, $R$ follows the relation, $\Omega D^{2} \sim A \sim R^{2}$ where $\Omega$ is the observed solid angle, $D$ is the distance, and $A$ is the area of detection. We also add an extreme case that could represent a smaller and denser HVC complex by simply cutting down the radius in the previous intermediate case to half so that the relation instead becomes $\Omega D^{2} \sim A \sim 2R^{2}$. Note that the mass of the HVC complex is the same in both intermediate and extreme density case and that the H{\small I} volume number density of the HVC complex ranges from $1.14 \times 10^{-5}~cm^{-3}$ (intermediate GCN) to $2.03 \times 10^{-2}~cm^{-3}$ (extreme WD). By considering two different velocities and two different sizes (one as the intermediate case and the other being extreme) for the approaching HVC complexes, we set up four different infall cases. A representative simulation run of the collision scenario between the disk and Complex C is shown in Figure \ref{fig:1}. The figure illustrates the most extreme case (higher velocity and density) among the four possible setups for Complex C where the vertical velocity and the H{\small I} volume number density are $143.9~ km~ s^{-1}$ and $5.20 \times 10^{-4}~cm^{-3}$, respectively.

\subsection{Estimating Fuel Efficiency}

\begin{figure}[h!]
	\centering
    \includegraphics[width=1.0\columnwidth]{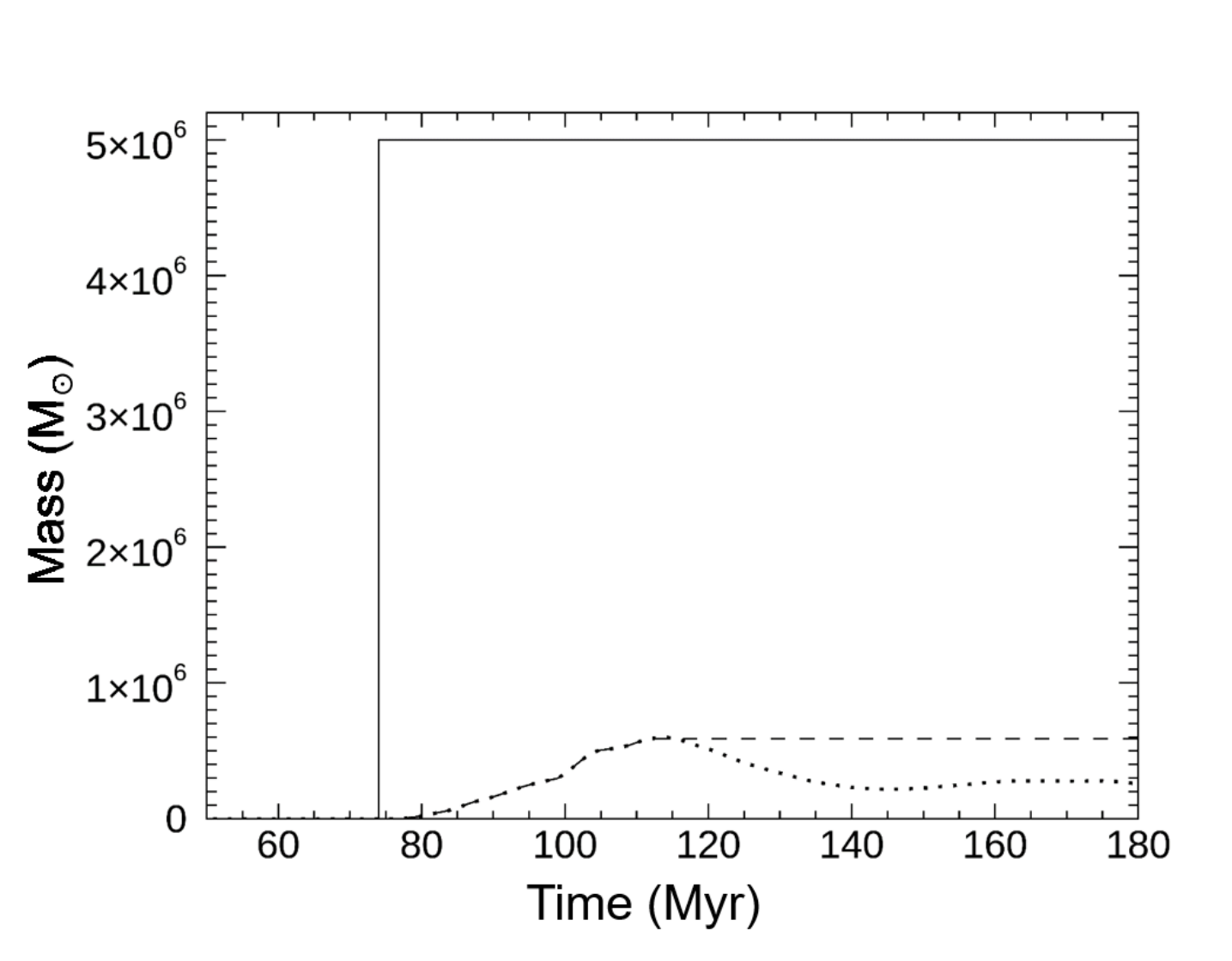}
    \caption{{\bf Time evolution of the mass infused into the disk from Complex C. The dotted line is calculated from the simulation results shown in Figure \ref{fig:1} and the dashed line shows how the missing long-term physical effects are compensated by assuming that the infusion of mass is completed after $2 \tau_{char}$ (112 Myr) since impact. The solid line illustrates the traditional estimation of mass infusion based upon the simple assumption that the mass of the infalling complex is fully accreted onto the disk.}}
    \label{fig:2}
\end{figure}

From the collision between the HVC complex and the disk, a shock is generated and the velocity of the forward shock which propagates through the galactic disk $v_{sd}$ is estimated from,
\begin{equation}
v_{sd} = \frac{4}{3} \frac{1}{1+\sqrt{\rho_{d}/\rho_{i}}}v_{i} ~,
\label{eq:2}
\end{equation}
where $\rho_{d}$ is the volume density of the galactic disk, $\rho_{i}$ is the volume density of the HVC complex, and $v_{i}$ is the velocity of the HVC complex \citep{tenta1981,lee1996,delvalle2018}. Without gravity and cooling included in our simulations, the rear-shock that is generated from the collision will be continuously expanding back into the ISM. Also some amount of the disk material would be permanently removed from its original position. Therefore, we define the characteristic time $\tau_{char}=h_d / v_{sd}$, where $h_d$ is the thickness of the disk in order to approximate the long-term evolution by assuming that the infusion process is completed at $t= t_i + 2 \tau_{char}$, where $t_i$ is the time at impact. For such time period, we investigate the time evolution of the HVC mass that is infused into the disk and apply equation (\ref{eq:1}) to further estimate the rate of mass infusion for multiple HVC collisions. For an example, the time evolution of the mass infused into the disk from the Complex C inflow is illustrated in Figure \ref{fig:2}. Note that this plot is made from the simulation run shown in Figure \ref{fig:1}. As full accretion is the underlying assumption in the traditional scheme, the fuel efficiency is estimated as the ratio of infused HVC mass obtained from the simulation results to the traditional prediction, i.e., the total HVC mass fully accreted onto the disk. Figure \ref{fig:2} shows that the fuel efficiency can be quite low because the HVC material cannot penetrate into the disk. This occurs mainly because the density of Complex C is lower than that of the disk. As illustrated in Figure \ref{fig:1},  the spherical complex is deformed into a horizontally flat shape and bounces back into the ISM when the low-density cloud impacts the high-density disk. In contrast, we find that the high-density cloud penetrates into (and passes through) the low-density disk although we do not show the results of our test simulations.


\begin{figure*}[t!]
	\centering
    \includegraphics[width=2.0\columnwidth]{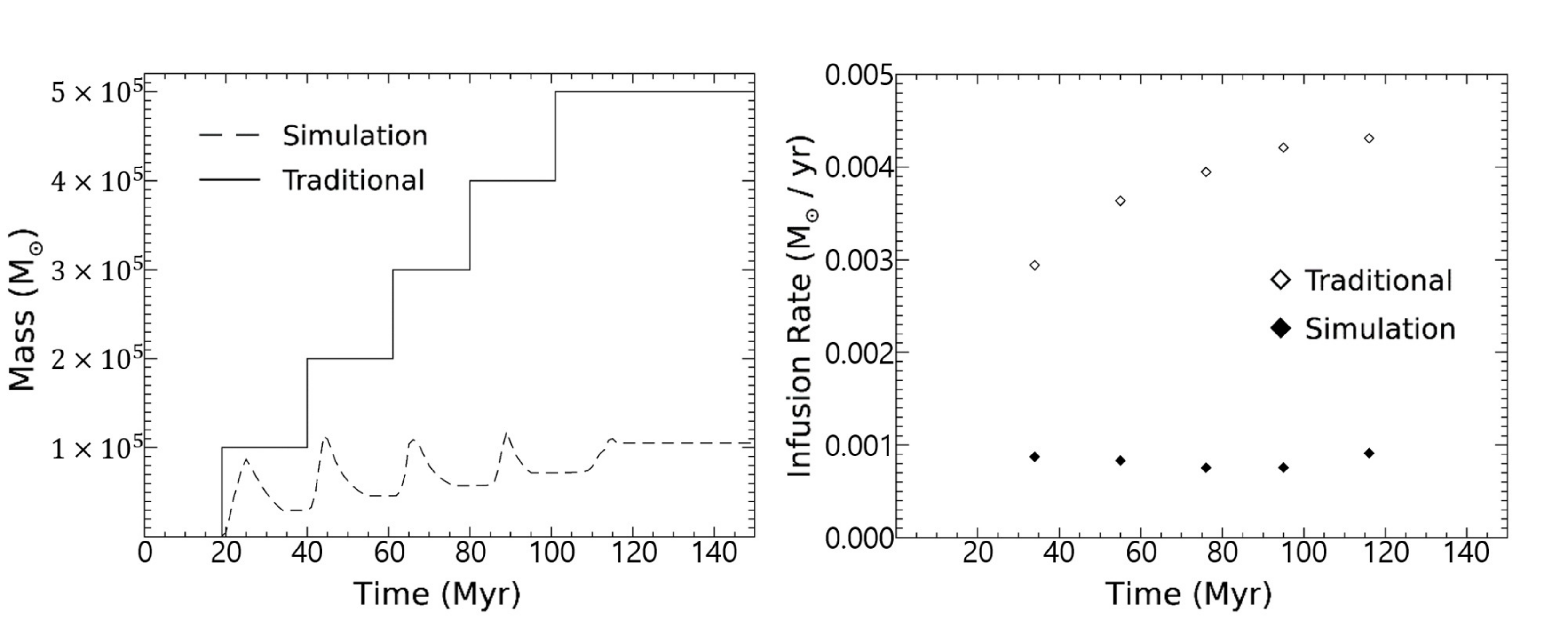}
    \caption{Five HVCs with the same mass ($1.0\times10^{5}M_{\sun}$) and velocity (100 $km~ s^{-1}$) but different distances (2, 4, 6, 8, and 10~kpc) are inflowing into the disk. The mass infusion into the disk as a function of time (left panel) and the infusion rate at time points when the infusion is completed for each HVC (right panel) are shown.  The solid line on the left panel represents the traditional interpretation of HVC mass accretion and the dotted line represents our simulation result. Similarly, empty and filled symbols on the right panel represent the mass infusion rate in the traditional scheme and from the simulation result, respectively.}
    \label{fig:3}
\end{figure*}

\begin{figure*}[t!]
	\centering
    \includegraphics[width=2.0\columnwidth]{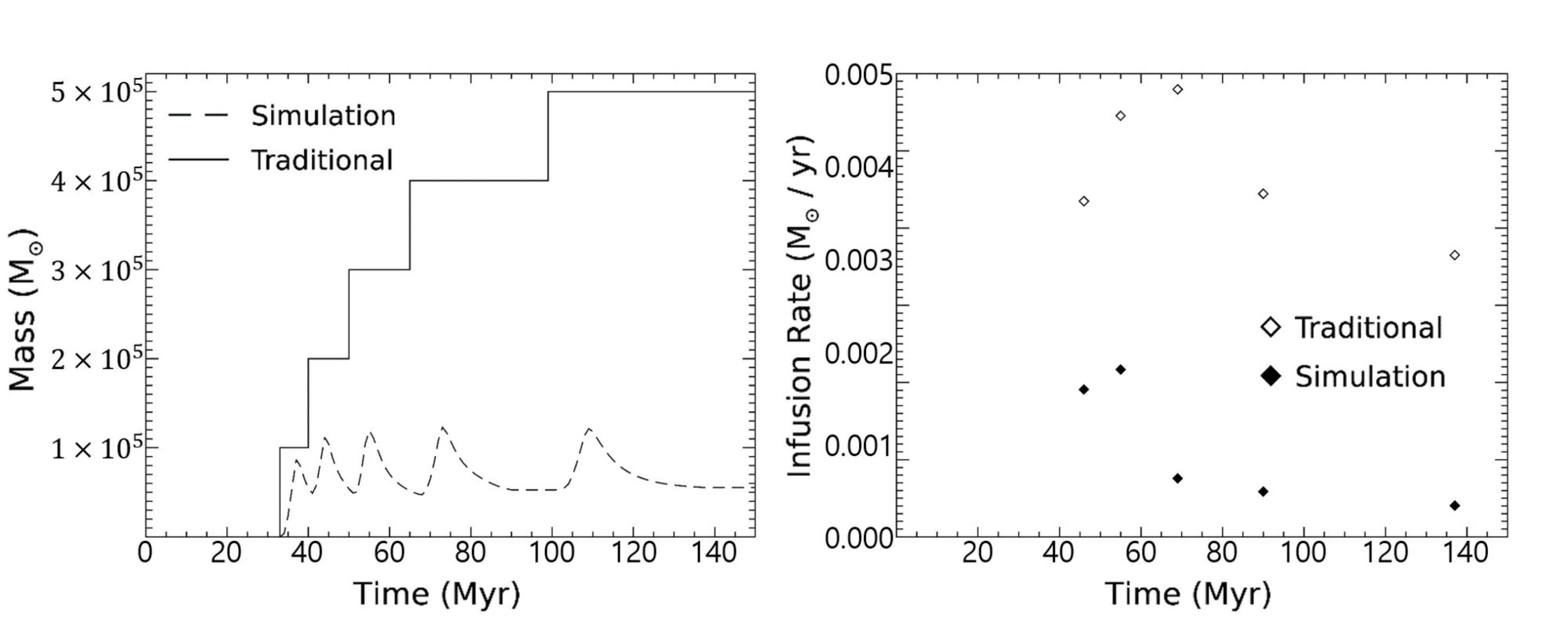}
    \caption{Five HVCs with the same mass ($1.0\times10^{5}M_{\sun}$) and distance (4 kpc) but different velocities (40, 60, 80, 100 and 120 $km~ s^{-1}$) are inflowing into the disk. The mass infusion into the disk as a function of time (left panel) and the infusion rate at time points when the infusion is completed for each HVC (right panel) are shown. The solid line on the left panel represents the traditional interpretation of HVC mass accretion and the dotted line represents our simulation result. Similarly, empty and filled symbols on the right panel represent the mass infusion rate in the traditional scheme and from the simulation result, respectively.}
    \label{fig:4}
\end{figure*}

\begin{figure*}[t!]
	\centering
    \includegraphics[width=2.0\columnwidth]{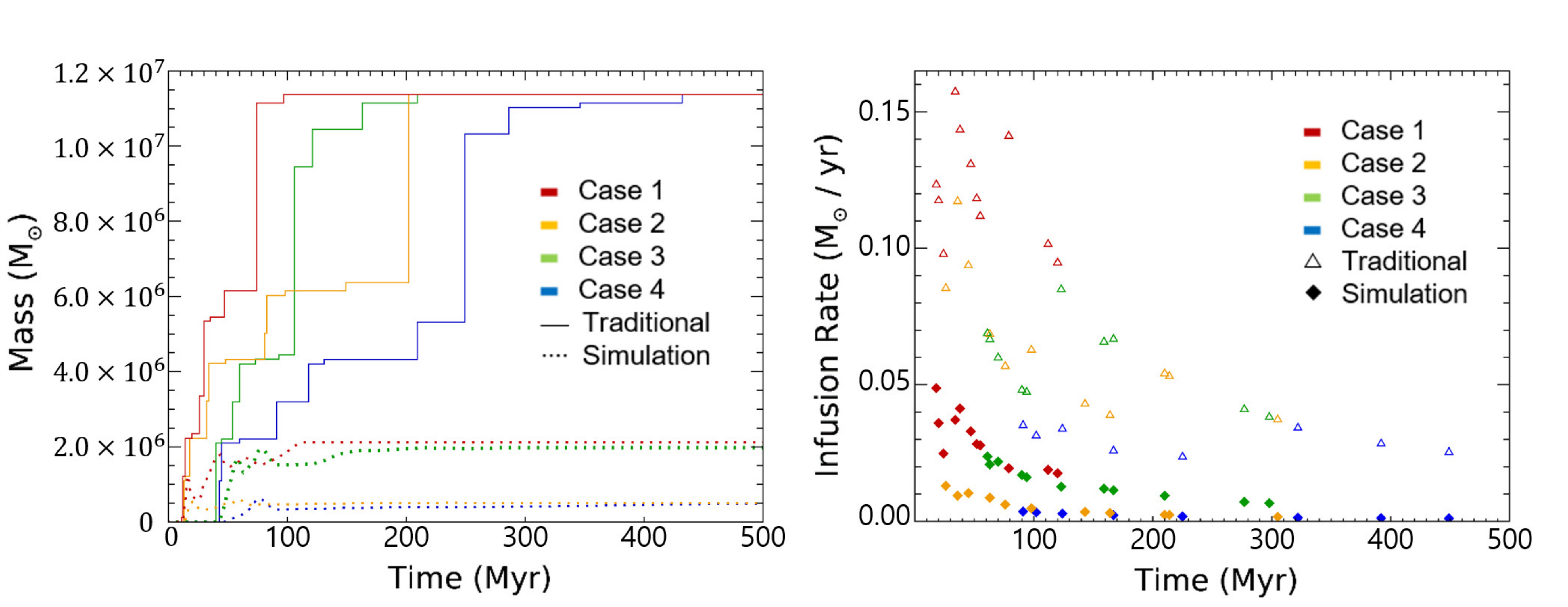}
    \caption{The mass infusion into the disk as a function of time (left panel) and the infusion rate at time points when the infusion process is completed for each incoming HVC (right panel). Solid lines on the left panel represent the traditional interpretation of HVC mass accretion while dotted lines our simulation results. Empty triangles and filled diamonds on the right panel represent the mass infusion rate from the traditional approach and from our simulation results, respectively. Total 11 HVC complexes are inflowing into the galactic disk and we consider 4 different infall cases that are represented with different colors (red, yellow, green, and blue). Case 1 is indicated in red with the HVCs having an extreme velocity and density. Case 2 is in yellow with extreme velocity but intermediate density. Case 3 is shown in green with intermediate velocity but extreme density. Case 4 is in blue with the complexes having both intermediate velocity and density.}
    \label{fig:5}
\end{figure*}

\section{Results} \label{sec:results}
\subsection{Proof of Concept} \label{subsec:proof}
We set up two different infall cases where five HVCs with the same mass of $1.0\times10^{5}M_{\sun}$ and the H{\small I} volume number density of $1.0 \times 10^{-2}~cm^{-3}$ fall down on the galactic disk. For the first case, the velocity is fixed at 100 $km~ s^{-1}$ but the distance of each HVC increases from 2 kpc to 10 kpc with an increment of 2 kpc. In the second case, the distance is fixed at 4 kpc but the velocity of the HVCs decreases from 120 $km~ s^{-1}$ to 40 $km~ s^{-1}$ with a decrement of 20 $km~s^{-1}$. As full and steady accretion of mass is assumed until the time to impact in the traditional scheme, the infusion of mass as a function of time will look similar to a staircase function \citep{putman2012} as in the left panels of Figures \ref{fig:3} and \ref{fig:4}. On the other hand, the simulation results which are represented with dotted lines on the same panels show peaks where the maximum infusion of each HVC has occurred at sometime after impact and then smooth down as the infusion is further processed and finally completed. The panels on the right in Figures \ref{fig:3} and \ref{fig:4} show the HVC mass infusion rate (i.e., infall rate) estimated from equation (\ref{eq:1}) at times where each HVC has completed the infusion process. The simulation result is represented with filled diamonds and the traditional estimation with empty diamonds. The fuel efficiency estimated as the ratio of infused HVC mass from our simulations to the total mass in the traditional estimation is 0.211 for the first case and 0.111 for the second. The average efficiency of the infall rate is 0.217 for the first case and 0.246 for the second. While detailed investigation on the interaction between an HVC complex and the galactic disk, which depends on the density, size, velocity, and distance of the complex is further required, we see in both cases that the infusion (or supply) rate of mass from HVCs into the galactic disk is overestimated in the traditional estimation compared to that in the simulation where we consider the interaction between HVCs and the disk. 

\subsection{HVC Complexes} \label{subsec:hvcs}
Provided the HVC parameters in Table \ref{tab:hvcinput}, we set up four different infall cases from the combinations between two different velocities and two different densities. The results for these cases are shown in Figure  \ref{fig:5}. While the HVC complexes in case 1 and case 2 share the same velocity, the ram pressure and drag force acting on the complexes are different due to different densities when moving through the halo. As a result, the complexes in case 2 generally have later impact times compared to those of the complexes in case 1 and this also applies between case 3 and case 4 which have the same intermediate velocities but different densities. As in the left panels of Figures \ref{fig:3} and \ref{fig:4}, the traditional estimation of the HVC mass infusion as a function of time is illustrated as a staircase function although it looks different in each case. The results obtained with the hydrodynamic simulations are represented in dotted lines and the fuel efficiency is 0.186, 0.044, 0.174, and 0.042 for cases 1, 2, 3, and 4, respectively. Despite the difference in velocity, the final efficiency in case 1 is similar to that in case 3, where the two cases share the same density. Such feature is also found between case 2 and case 4. Since the complexes in cases 1 and 3 have a larger density than those in cases 2 and 4, we can conclude that the final efficiency depends strongly on the density than on the velocity of the cloud or it increases as the density of cloud increases when the hydrodynamical interaction between the cloud and disk is considered. This occurs likely because the densities of the HVC complexes in consideration are smaller than the density of the disk, but the denser complex still penetrates into the disk more efficiently. The right panel in Figure \ref{fig:5} shows the HVC mass infusion rate (i.e., infall rate) estimated from equation (\ref{eq:1}) at time points where the infusion process has been completed for each HVC complex. The average efficiency of the infusion rate is 0.249, 0.092, 0.247, and 0.072 for cases 1, 2, 3, and 4, respectively. Again, the average efficiencies of the infall rate in cases 1 and 3 (2 and 4) are similar regardless of their velocity differences, but efficiencies in cases 1 and 3, where the density of the complexes are larger, are higher than those in cases 2 and 4. Overall, we see in all 4 cases that the supplied amount of mass from infalling HVC complexes into the galactic disk are overestimated in the traditional estimation compared to our simulation results that include the hydrodynamic interaction between HVCs and the disk. This is mainly because all the complexes in consideration have a lower density compared to the gaseous disk.  In other words, individual complexes, regardless of the selection of density and velocity, have a similar trend of fuel efficiency to that shown in Figure \ref{fig:2}.

\section{Discussion}
Though we have attempted to eliminate the long term dynamical effects during the collision between the galactic disk and an HVC complex by assuming that the infusion of mass from HVCs are completed at the time $2 \tau_{char}$ after impact, we are well aware of the fact that the simulation in this study is a simplified model and has limitations as we did not include gravity, cooling, and magnetic fields that could also play important roles in the interaction between the disk and HVC. Below we give a list of improvements that could be made on our current simulation setup and briefly discuss how each element could act on the fuel efficiency.

1. Adopting a non-uniform density profile for the HVC complexes, galactic disk, and halo material:
\noindent It is shown in Section \ref{subsec:hvcs} that the fuel efficiency depends more strongly on the density of the HVC complexes than on the velocity. Further considering the case where the complexes have a non-uniform density profile, the lower-density gas at the edge of the complex will mix with the ISM in the halo instead of being infused into the disk due to shear instabilities \citep[e.g.,][]{kwak2011} which in turn will result in lower fuel efficiency\footnote{Interaction of the infalling HVCs with the gas pushed up into the halo by stellar and/or supernova feedback would also decrease the fuel supply rate \citep{FraternaliBinney2008}.}.

2. Considering the Galactic gravitational field:
\noindent Gravity is a fundamental physical process in almost every astrophysical situation. However, at reasonably short time scales of the infusion process, the hydrodynamic interactions are dominant over the effects of gravity. Therefore we expect the impact on fuel efficiency from gravitational fields would be more significant at later times long after the collision between the complex and the disk. We eliminated the gravitational effects by assuming that the infusion process is completed at the time $2 \tau_{char}$ after impact. However, if gravity were considered instead of making such an assumption, there could be cases of steady accretion when the cloud does not survive after impact or maybe the fuel efficiency will increase in general as the HVC materials would ``tend" to stick better onto the galactic disk. We should also consider that the velocity of the HVC complexes would increase at the time of impact due to gravitational acceleration.

3. Taking into account the magnetic fields in the Galaxy:
\noindent How the magnetic field affects fuel efficiency can be complicated as it depends on the orientation of the field. In general, the magnetic field will restrain the growth of shear instabilities \citep{maclow1994,jones1996}, which could increase the fuel efficiency by making the HVCs survive longer in the halo. However, the field lines will be compressed during the collision when their orientation is perpendicular to the HVC velocity. In this case, the complex would encounter resistance which would potentially prevent the infusion of mass and lower the fuel efficiency \citep{santillan1999,kwak2009}. 

4. Considering cooling effects during the infusion process:
\noindent Since there is no cooling included in our simulation, the rear-shock will be ever expanding back into the ISM. Assuming that radiative cooling is important, the HVC material bounced back from the disk will mix with the ISM close to the disk instead of propagating far away from the galactic disk. From then gravity could take on and allow additional accretion which will increase the fuel efficiency in the long term.

5. Considering the actual path of HVC inflow (i.e., non-vertical drop):
\noindent The early hydrodynamic interaction timescale will increase when the path of HVC inflow is non-vertical. The magnetic field orientation along with other physical conditions of the inflowing HVCs must be considered, however, a longer mixing-time would generally allow more accretion of mass from inflow and increase the fuel efficiency.

\section{Conclusion}
This work is motivated from a simple idea that HVC complexes may not continuously nor fully accrete on the galactic disk since the hydrodynamic interaction between the disk and HVCs is not considered in the traditional method of estimating the infall. We selected 11 HVC complexes and constructed 4 infall cases to point out that the infall rate from the traditional approach could be an overestimated value and our simulation results show that the efficiency of the infusion rate from HVC inflow could be as low as $\sim 0.072$ when compared to the full accretion case from the traditional approach. The fuel efficiency is low because the densities of the complexes are lower than the density of the disk, and thus the complexes in consideration do not penetrate into the disk. However, we find that the density of the HVC complex affects the fuel efficiency and that its effect is more significant than that of the velocity. The fuel efficiency increases as the density of the HVC complex increases when the hydrodynamical interaction between the disk and HVC is considered.

Limitations of the current study come from adopting a uniform density profile for the HVC complexes, galactic disk, and ISM and also missing some potentially important physical processes such as gravity, magnetic fields, and cooling in our simulations. However, we expect the mass accretion from HVC infall to be still inefficient compared to the traditional prediction even when such physical processes are included.

Though our next task is to better investigate the true fuel supply rate from HVC infall by improving the current limitations like adopting a non-uniform density profile for the gaseous objects and adding the physical processes that were neglected in the current study, the ultimate goal of studying the effects of HVCs in the GCE is to figure out how stars form from HVC infall, which also requires a detailed investigation on the amount of H$_{2}$ that is converted from the H{\small I} which has survived from HVC infall.

\acknowledgements
This work has been supported by NRF (National Research Foundation of Korea) Grant funded by the Korean Ministry of Education (NRF-2015H1A2A1031629-Global Ph.D. Fellowship Program). KK was also supported by Basic Science Research Program through the National Research Foundation of Korea (Grant No. 2016R1A5A1013277, 2016R1D1A1B03936169).



\end{document}